\documentclass{article}

\PassOptionsToPackage{numbers}{natbib}
\usepackage[final]{neurips_2021}

\usepackage[utf8]{inputenc} 
\usepackage[T1]{fontenc}    
\usepackage{hyperref}       
\usepackage{url}            
\usepackage{booktabs}       
\usepackage{amsfonts}       
\usepackage{nicefrac}       
\usepackage{microtype}      
\usepackage{xcolor}         

\title{Could AI Democratise Education? \\ Socio-Technical Imaginaries of an EdTech Revolution}

\author{
Sahan Bulathwela\thanks{Equal Contribution}\\
    Centre for Artificial Intelligence,\\University College London, UK\\
    \texttt{m.bulathwela@ucl.ac.uk} \\ 
\And  Mar\'ia P\'erez-Ortiz$^*$\\
    Centre for Artificial Intelligence,\\ University College London, UK\\
    \texttt{maria.perez@ucl.ac.uk} \\
\And Catherine Holloway\\
    UCL Interaction Centre \&\\Global Disability Innovation Hub,\\ London, UK\\
    \texttt{c.holloway@ucl.ac.uk} \\
\And John Shawe-Taylor\\  
    Centre for Artificial Intelligence,\\ University College London, UK\\ 
    \texttt{j.shawe-taylor@ucl.ac.uk}
  }

\begin{document}

\maketitle

\begin{abstract}
Artificial Intelligence (AI) in Education has been said to have the potential for building more personalised curricula, as well as democratising education worldwide and creating a Renaissance of new ways of teaching and learning. Millions of students are already starting to benefit from the use of these technologies, but millions more around the world are not. If this trend continues, the first delivery of AI in Education could be greater educational inequality, along with a global misallocation of educational resources motivated by the current technological determinism narrative. In this paper, we focus on speculating and posing questions around the future of AI in Education, with the aim of starting the pressing conversation that would set the right foundations for the new generation of education that is permeated by technology. This paper starts by synthesising how AI might change how we learn and teach, focusing specifically on the case of personalised learning companions, and then move to discuss some socio-technical features that will be crucial for avoiding the perils of these AI systems worldwide (and perhaps ensuring their success). This paper also discusses the potential of using AI together with free, participatory and democratic resources, such as Wikipedia, Open Educational Resources and open-source tools. We also emphasise the need for collectively designing human-centered, transparent, interactive and collaborative AI-based algorithms that empower and give complete agency to stakeholders, as well as support new emerging pedagogies. Finally, we ask what would it take for this educational revolution to provide egalitarian and empowering access to education, beyond any political, cultural, language, geographical and learning ability barriers.
\end{abstract}

\section{Introduction}
Quality education is essential for achieving a sustainable world, benefiting individuals and societies alike. A strong educational system can broaden access to opportunities, improve health, bolster the resilience of communities and institutions, drive long-term economic growth, reduce poverty and spur innovation \cite{nilsson2016policy}. Education could also bring a fundamental shift in how we think, act, and relate to our responsibilities toward one another and the planet, helping to nurture a new generation that supports the transition to a prosperous and sustainable future.
In this sense, we believe we are all stakeholders of education systems. Firstly because we are all learners, but also because of the global benefits that an empowering and strong educational system could bring. Thus, we argue that stakeholders of education go beyond learners and teachers, and include administrators, policy makers, parents and other groups that play a role in this complex system as well as benefit from it.
While countries have significantly increased access to education, being in school is not the same as learning \cite{world2018world}. 
Worldwide, hundreds of millions of children still reach adulthood without even the most basic skills \cite{united2020global}. In this paper, we discuss the promise and the peril of Artificial Intelligence (AI) in Education, emphasising the need for a critical socio-technical design accompanied by open educational resources and tools. 

AI is changing the  skills needed in our
global and innovation centred
world and
exemplifying
novel methods of
teaching and learning \cite{holmes2019artificial}. 
AI in Education (AIEd) \cite{holmes2019artificial} covers, among others, intelligent, personalised and conversational educative systems (e.g. systems which provide scoring, assessment, feedback and hints or that match users for collaborative learning) with the aim to support stakeholders and put them in control of the learning process and design.
One of the most ambitious use cases of AIEd, Intelligent Tutoring Systems (ITS), has shown experimentally to lead to similar learning gains than face-to-face one-on-one instruction \cite{woolf2010},
while personalised learning, in general, can improve learning gains of an average student in the order of two standard deviations \cite{Corbett2001}.


Despite this immense theoretical potential of personalised technology-enhanced education and the large amount of monetary investment in these learning technologies, such innovations have not delivered much practical results yet \cite{its_press_release}.
Less ambitious AIEd technologies are, however, slowly starting to play a role in providing support to teachers, students, and the learning process more broadly \cite{x5gon,goal_based_edrec}. But without a doubt, not all are benefiting equally \cite{holmes2021ai}. 
 Historically, technological change has been shown to increase between-country inequality\footnote{\url{https://unctad.org/page/technology-and-innovation-report-2021}}. Similarly, AIEd could bring risks of exacerbating the wide education and opportunity 
 gap, 
which at the same time will have other negative consequences globally such as an imbalance of wealth/power concentration and human capital flight.

This work asks and imagines how could this AI-based educational potential benefit everyone equally. 
In our view, providing high quality education to almost eight billion people requires socio-technical imaginaries (i.e. future-oriented visions of connected social and technological orders), specially of ways to scale education, which is the focus of this work.
 For example, the recent boom of Open Educational Resources \cite{unesco1} 
 could mean that the democratisation of learning may be within reach, but only if the objectives of AIEd are adequately aligned and collectively designed, with accessibility, equity, 
 openness
 and inclusion at their core.  
This paper starts summarising how AI could support the task of learners and teachers at scale, reflecting on some of the social and technical challenges. We then aim to identify some of the barriers to benefactor everyone with decades of advances in AIEd, with the aim of starting the dialogue to ensure these tools narrow the educational inequity gap rather than increasing it. 
We finally discuss some promising resources that could be used to leverage this participatory education revolution. Ultimately, this paper aims to pose the question: \emph{Could a AI tools help us to democratise education? What would it take for that to happen?}



\section{AI in Education (AIEd): The Promise and the Peril}

 \paragraph{The promise} AIEd has been under active research for several decades now \cite{corbett1994knowledge,peek_orsum} and has made significant advances in different fronts, specially in areas such as intelligent tutoring (ITS) \cite{Yudelson13}, educational recommendations \cite{goal_based_edrec}, MOOC management \cite{ramesh2014learning,Guo_vid_prod} and taxonomy/prerequisite detection \cite{yu2020mooccube}, among others. 
Recent work has also shown that AI could be used to accumulate learning resources at scale \cite{x5gon}, as well as enrich these to break language barriers by creating cross-lingual translations/transcriptions \cite{baquero-arnal-etal-2019-mllp}, domain/language-agnostic topic annotations \cite{wikifier} and visual and interactive content summaries that support the learner \cite{x5learn,10.1145/3397482.3450721}.
The ultimate ambition of AIEd would be a lifelong learning companion
, that understands the strengths and weaknesses of individual learners to present materials and exercises to increase their learning gains, while providing prompt feedback when needed. Being able to cooperatively operate across languages, cultures and special needs of individuals would make this companion humane. Above all, this companion should interplay with political and operational constraints, respecting privacy, safety and prioritising the learner autonomy and agency, moving away from prescriptive AI. Such technology should also allow teachers to use their training and experience to fulfil less mundane tasks such as personal attention, advance pedagogy, pastoral care and other complex support tasks that preserve equity in the classroom. These systems can also enable teachers, educators and researchers to carry out valuable research and hypothesis testing that will lead to a better understanding of pedagogy itself. While being an ambitious and impactful destination (dreamed since 1972 \cite{its_press_release}), achieving such a technology may be further down the line than we think. 
We believe that by shifting the paradigm to \emph{augmenting} teacher/stakeholder capabilities, rather than replacing them, AIEd can bear more fruits in the short-term, while journeying at the same time towards a sustainable large-scale lifelong educational practice enhanced by AI.

 \paragraph{The peril (and the challenges)}
 We must keep in mind
that increasing access to education remains predominantly
a political and social issue \cite{holmes2021ai} and that disconnection between a technology and the surroundings where it operates leads to consistent failures \cite{toyama2010can}. AI technologies could help education in different ways, but
they are unlikely to offer a solution on their own.
 Even more, if AIEd technologies are not designed appropriately and collaboratively and deployed with the right infrastructure and across nations, we believe that one of the greatest perils we could face is for these to exacerbate educational inequality in the world (which could itself have many other consequences on a global scale), as well as divert educational resources that could be put to more effective uses and propagate dangerous biases of different nature at scale \cite{world2018world}. 
  Access to these technologies is also about power. After all, in a society that is thoroughly permeated by technology, those who possess access to it, can influence processes and will have greater opportunities.
Various recent works have highlighted other challenges to circumvent before benefiting from the opportunities of AIEd. For example, IDIA\footnote{\url{https://www.idiainnovation.org}} \cite{idia_report} identifies several technical challenges general to AI: availability of quality data, accountability, transparency and addressing biases.  The recent UNESCO\footnote{\url{https://en.unesco.org}} \cite{holmes2021ai} and IRCAI\footnote{\url{https://ircai.org}} \cite{ircai_report} reports narrow down the challenges in AIEd to scalable content understanding and fact checking, learner modelling, personalised learning, transparency and scrutiny. These reports also expand on scalable evaluation, which is an essential part of verifying learning \cite{Yudelson13}, and the need for addressing lifelong learning \cite{trueeducation}.  

Amid the promises of technology-enhanced education, the success stories seldom benefit developing nations. To start with, there is a geographical, cultural and language imbalance in terms of open education repositories around the world \cite{oer_worldmap}. Even more, the majority of educational materials that have been accumulated via AI 
are in popular European languages, due to the narrow selection of translation models that are of current focus in the research community \cite{x5gon}. 
Even among available translation models, the performance only marginally surpass "humanly-acceptable" level \cite{x5gon_d5_3}. This is far from the needed quality for learning purposes, where translation and transcription errors can easily impair the learning experience. The scientific community is also focused on introducing learner models, ITS solutions and knowledge taxonomies that are very specific to narrowly scoped datasets and technologies, giving much less emphasis to low-resource settings \cite{assist_math1,yu2020mooccube,choi2020ednet}. 
Apart from all these shortcomings of existing research, there is a whole realm of considerations such as Internet connectivity, unequal access to digital devices
as well as accessibility needs that need to be accommodated when innovating responsibly in this space. 
For example, 80\% of the world’s population of people with disability
live in developing countries. With lack of educational resources, these communities struggle to gain the skills necessary to create a livelihood, contributing to the known link between poverty and disability \cite{holloway2019disability}. Assistive technology, with the help of domain clinicians can help this community and AI can expedite this process \cite{gdi_report} (e.g. Google Euphonia). 


\section{Proposed Pillars for AIEd}

With the ambitious objective of building and maintaining a sustainable large-scale AIEd ecosystem that facilitates equitable, high-quality lifelong learning opportunities for all, we identify several essential needs and promising tools: (1) \emph{Open Education}: A large growing collection of freely available and accessible educational resources 
    with appropriate diversity to suit a global learner population.
    (2) A \emph{unifying taxonomy of knowledge}: A language-agnostic representation of knowledge that can be used to build AIEd tools (we propose Wikipedia).
    (3) \emph{Human-centred AI}: A suite of fair, interactive, collaborative and transparent AI algorithms that give full agency to the end-user.

    Apart from these three pillars, we also emphasize the need for \emph{open-source} to promote community engagement and  \emph{critical analysis documentation}.
    We believe that both open education and open-source are essential 
foundations for enabling civic engagement with 
technologies \cite{voigt2021open}. These could empower and emancipate communities by not only allowing them to easily adopt such tools and resources, but also enable them to change them to their own needs and participate in their design and  business models, improving independent experimentation and locally situated economies. 
We hope that by designing together as a community these tools we can set the goals and fundamental basis right, while at the same time by making them open-source we can give rise to resilient civic societies that are largely independent of the global infrastructures and can tap into their own local resources and knowledge \cite{voigt2021open}.
Moreover, AI systems are fundamentally
socio-technical \cite{dignum2021role}, including the social context in which they are developed, used and
acted upon. 
The  processes  by  which  systems  are  developed  entail  a  long  list  of  decisions  
by  designers,  developers  and   stakeholders,  many  of  them  of   societal,  legal  
or ethical nature \cite{dignum2021role}.
Thus, AIEd solutions should be accompanied with critical documentation \cite{datasheets} that states the design rationale, as well as any limitations of the educational datasets/tools/resources and the context in which they were developed. 


\subsection{Open Educational Resources (OER)}

Identifying critical barriers surrounding access, quality and costs of information and knowledge available over the Internet, the OER initiative was founded to improve global access to knowledge \cite{hylen2021open}.
OERs are open licensed educational materials distributed on the Internet with significantly less restrictions than other educational materials, enabling others to retain, reuse, revise, remix and redistribute content \cite{unesco1}. This innovation succeeds in scaling rapidly by providing a toolkit that minimises the effort of creating teaching materials from scratch (e.g. through innovative licensing schemes and aggressive growth models such as the content explosion model \cite{pawlowski2007open} and the open educational practice \cite{TVET2018}). 
The success of this innovation also stem from the community's tendency to constantly facilitate design hackathons \cite{curriculum_report,paris_mlw_report} that connect designers, educational practitioners, developers and other stakeholders to sit in the same table to develop solutions. 
The true potential of OERs is only starting to show tangibility, with OER collections accumulating more than 100,000s materials, 
these having been curated \cite{quality_features, context_agnostic_engagement}, translated to multiple languages using AI, as well as transcribed and annotated \cite{x5gon}.
With the use of support tools such as interactive translation systems \cite{kreutzercorrect}, the cross-lingual translation educational materials can be expedited making many rich educational resources available to diverse communities in their native tongues. 
The feasibility of enriching these materials and presenting them to users in an intelligent user interface has been demonstrated recently \cite{x5learn}.
 Furthermore, accessible formats or templates for creating accessible content for all learners, can be built in from the start. In our view, the OER initiative represents the most promising cross-domain culturally-diverse collection of materials for democratising education.

\subsection{Wikipedia as a Standardised Knowledge Base}


The need for a unifying knowledge base and taxonomy has been one of the greatest challenges faced by the AIEd community since its inception. 
This is, for example, essential for deploying AIEd systems at large scale, as these solutions can not be handcrafted. Such systems will need to understand the universal structure and direction of knowledge, identify automatically knowledge prerequisites, topics covered and difficulty of educational materials, while at the same time filter them by metrics of quality assurance, with (for example) the goal of matching the most appropriate materials to learners. All of this needs to be achieved across multiple modalities of knowledge, languages and cultures. 

Wikipedia remains the world's largest and most up-to-date Encyclopedia. It achieves this i) by using technologies that support humans to contribute information 
and ii) by including crowdsourcing at the heart of every aspect of Wikipedia.
Wikipedia also leverages AI to help scale this human information management operation, for example augmenting intelligence in article quality assessment \cite{wiki_wang}, defending contributors from abuse \cite{wiki_troll} and various other tasks \cite{forbes_ai}. 
We envision multiple opportunities in utilising Wikipedia to create educational tools that support equitable lifelong learning opportunities for all. First, as a universal knowledge base, Wikipedia can become the common taxonomy enabling inter-operability among different educational standards and materials that belong to different nations and educational systems. As an example, many governments and organisations have invested resources and expertise on developing curricula, taxonomies, teaching guidelines and learning resources that uplift the quality of education in local contexts. However, this localisation has posed grievous challenges in cross compatibility of knowledge \cite{curriculum_report}, as it can not be reused easily. 
Using novel entity linking approaches, there is opportunity now to ground curricula that originate from different systems into a single taxonomy allowing the global population to discover relevant educational materials that are enriched beyond their local environments. The utility of such a global taxonomy has already been shown in social media \cite{10.1145/2993318.2993332}, and educational recommenders \cite{truelearn}. Secondly, having a humanly-intuitive taxonomy  in its foundation (as Wikipedia does) also allows embedding expert supervision and scrutiny into the process \cite{bull_trust,molnar2020interpretable}. Finally, such grounding opens up avenues to cross-disciplinary lifelong learning experiences (across time, language and geography) as the global taxonomy is domain agnostic and convertible to local taxonomies. 
Wikipedia, as an open platform, also comes with its weaknesses, such as exposure to social biases and challenges in fact-checking. However, by acknowledging and identifying such weaknesses, we can work towards mitigating them and engage with stakeholders to uplift the quality of this living taxonomy.

\subsection{Human-Centred AI-based Educational Tools}
When deploying AIEd, we need to consider the prevalent learnings from AI for social good and developing nations research \cite{tomavsev2020ai}. One core question we need to ask is if we are really accounting for the technological, societal and cultural differences across nations. Are we ensuring that the
interests of low and middle income countries are represented
in key debates and decisions? Are we creating the necessary bridges between
these nations and countries where the AI is currently being developed?
Currently, the majority of AI solutions we create revolve around \emph{ideal} scenarios defined by datasets that are created in controlled environments in the developed world. First of all, it is crucial that we ensure that our test beds are not far from real life imperfect settings where these solutions may operate in, and that we engage everyone involved in the design process.

Because of this, we believe that AI-based educational tools should embrace design patterns within the umbrella of human-centred machine learning. Specifically, in this framework we require both humans and machines simultaneously in the loop. Usually, we would require humans to give feedback to the algorithms to learn. This could be a mix of explicit and implicit feedback, so models can improve over time \cite{trueeducation}. Most specifically, allowing for hybrid human-machine interaction and collaboration is of crucial importance in such an educative system, to give the stakeholder full control and access to manipulate their own personalised learning tools and put the tools at their complete service.
For us, this means that the user needs to be able to design up to some extent their own tools, as well as interact, query and change the model’s perception of them, and explicitly indicate timely preferences, needs, and goals, which can guide the tools. 
This is, AIEd needs to move from prescriptive algorithms to collaboration with the human. 
Part of these requirements not only apply to AI but more generally to the design of 
the human-computer interaction. 
We believe that rather than predefined curricula or behaviour change, we need to offer dynamic 
paths for users to choose. Enabling the user to make informed decisions and allowing a mutually productive dialogue between the human and the machine should be the primary goal of AIEd. This is in contrast to the prevalent use of recommender systems, whereby suggestions are
presented in one direction from the system to the user. 
Transparency and privacy should also be key to such systems, allowing stakeholders to understand the potential and limitations of these algorithms and to decide on what data should the algorithms store and use for reasoning.  
We hope that trust from the user can be achieved by demonstrating transparency and integrity in all regards, and by steering clear of any business models that might compromise the trust relationship with users \cite{bull_trust}.
The machine needs to take a full supporting role, with users ultimately deciding when and how to use the tools. 
This could mean that humanly-interpretable latent variable models should be preferred over their black-box counterparts \cite{trueeducation}. 

We believe that a pivotal role of AIEd (and an ambitious challenge) will be to move towards innovative emerging pedagogies,
such as formative analytics (where learners are provided with information for self-regulated learning), teach back (providing learners an opportunity to teach their learnings) and learning with robots  (where repetitive tasks such as assessment and hint giving are automated to free up teachers for cognitively demanding tasks) \cite{emerging_pedagogy}. With the right adaptation of AI-based education, these novel pedagogies could unleash great potential at a global scale. Other futuristic pedagogies, such as place-based learning \cite{kukulska2015mobile} and citizen enquiry \cite{curtis2018online}, which revolve around engaging learners in participating in an active problem-solving environment, could also foster great potential benefits.

\section{Discussion}

AI will impact
 education greatly.
 However, 
 virtually no research has been
undertaken, no guidelines have been agreed, no policies
have been developed, and no regulations have been enacted to
address the use of AI in education \cite{holmes2021ai}. It is time that we decide collectively what technology-enhanced education should mean, with the end goal of increasing access to high quality education for all. At the moment, the field is devoting much attention to personalised intelligent tutors. However, such ambitious use cases,  essentially proposing to replace teachers, are far from delivering any impactful and real-world results.  
We argue that massively investing resources in such technology poses the risk of misallocating necessary resources to enhance the current lack of access to high-quality education around the globe. 
 We argue that the community should focus instead on other low-hanging fruits, such as tools that support the role of teachers, while at the same time build a strong basis (in socio-technical terms) that could, in the future, support personalised intelligent tutors research. 
 
Perhaps if correctly designed and deployed, AI tools could deliver in the long run on their potential of: i) providing at scale empowering access to education beyond any political, cultural, language, geographical and learning ability barriers; ii) helping us create fulfilling, equitable and inclusive lifelong learning schools of the future; and iii) leveraging the so-called Reinassance of new ways of teaching and learning. 
However, like with any other technologies, the greatest challenge is how to design them to be a driver of equity and inclusion and not a source of greater inequality of opportunity.
 
  There are enormous technical, social, political and
 pedagogical challenges ahead for the field of AIEd.
 These are issues related to data and algorithms, on
pedagogical choices, on inclusion and the ‘digital divide’,
on children’s right to privacy, liberty and unhindered
development, and on equity in terms of gender, disability,
social and economic status, ethnic and cultural background,
and geographic location \cite{holmes2021ai}. 
 Our aim was to discuss a subset of these, with the hope of starting this dialogue and collectively design a global education revolution that will help us solve educational inequity. We propose a socio-technical solution to meet part of these challenges. This is, i) working together on developing and leveraging the power of language and culturally diverse open educational resources, which can be reused and consumed around the globe; ii) building standardised taxonomies and ontologies of knowledge, one of the greatest technical current challenges for AIEd; iii) investing in open-source technology to enable civic engagement with the design of these technologies and thus support their sustainable use in local communities; iv) prioritising research in transparent, scrutinisable, interactive and collaborative AI algorithms; and v) engagement in critical thinking and policy making, where we question the social norms and politics embedded in AIEd systems and direct technological change towards meeting societal needs and reducing inequalities.

 Before committing to a future where AI pervades learning, educationalists and
technologists need to guide society and governments to understand the potential
social  and  ethical  implications  of  this  technology \cite{dignum2021role}. 
 Engaging in speculation when designing technology is of the most crucial importance. We provide now a non-exhaustive list of question examples we believe the field of AIEd should be asking: Regarding technical terms, we could ask: 
 What is the nature of knowledge,
and how can it be represented and captured with AI? How could an AI system automatically gather, understand and filter educational resources suitable for each learner needs? How can AI help bridge the education gap for learners with disabilities?
How can AI-based tools prioritise transparency, keeping the human in the loop to support users to self-reflect on their learning path and give them agency?
 What learning metrics, if any, should guide these learning tools and algorithms? 
 How can personalisation algorithms support our diversity as individuals and communities?
Regarding the social aspect, we could ask: How could we scrutinise any biases as well as social prejudices embedded in these techniques? How can we collectively identify and document the limitations in the educational datasets/tools/resources? 
How can we support our communities to engage with the design of educational tools that concern us all? How can we all learn to identify faulty educational tool designs that plague the most vulnerable and collectively correct these? How could we successfully and safely iteratively prototype and evaluate these tools in the wild?  
Regarding pedagogy, examples could be:
 How can we promote design thinking in educative settings, where we allow the "what's wrong?" drive our pursuit of "what if"? 
 Could, ultimately, these AIEd systems engage communities to transform individuals, communities and the environment?
 How can such a system encourage critical and emerging pedagogies? 
Ultimately, the question we, as a research collective in AIEd, need to ask (and work towards) is: What would it take for AI to help us democratise quality education for all? Our claim is that AI on its own can not offer a solution, and we should, at the same time that we develop technology-enhanced educational tools,  address the political and social issues behind the unequal access to high-quality education. 

\section*{Acknowledgements}
This work is partially supported by the European Commission funded project "Humane AI: Toward AI Systems That Augment and Empower Humans by Understanding Us, our Society and the World Around Us" (grant 820437) and under the AT2030 Programme. The AT2030 programme is funded by the UK Aid from the UK government and led by the Global Disability Innovation Hub. We also thank Yvonne Rogers from UCL Interaction Centre and Davor Orlic from Knowledge 4 All Foundation for the valuable feedback provided during the discussions.

\bibliographystyle{plain}
{
\small
\bibliography{main}

\begin{thebibliography}{10}

\bibitem{bull_trust}
Norasnita Ahmad and Susan Bull.
\newblock Learner trust in learner model externalisations.
\newblock In {\em Proc.~of conf. on Artificial Intelligence in Education},
  2009.

\bibitem{world2018world}
World Bank.
\newblock {\em World development report 2018: Learning to realize education's
  promise}.
\newblock The World Bank, 2018.

\bibitem{baquero-arnal-etal-2019-mllp}
Pau Baquero-Arnal, Javier Iranzo-S{\'a}nchez, Jorge Civera, and Alfons Juan.
\newblock The {MLLP}-{UPV} {S}panish-{P}ortuguese and {P}ortuguese-{S}panish
  machine translation systems for {WMT}19 similar language translation task.
\newblock In {\em Proceedings of the Fourth Conference on Machine Translation
  (Volume 3: Shared Task Papers, Day 2)}, pages 179--184, Florence, Italy,
  August 2019. Association for Computational Linguistics.

\bibitem{wikifier}
Janez Brank, Gregor Leban, and Marko Grobelnik.
\newblock Annotating documents with relevant wikipedia concepts.
\newblock In {\em Proc.~of Slovenian KDD Conf. on Data Mining and Data
  Warehouses (SiKDD)}, 2017.

\bibitem{x5learn}
Sahan Bulathwela, Stefan Kreitmayer, and Mar\'{\i}a P\'{e}rez-Ortiz.
\newblock What's in it for me? augmenting recommended learning resources with
  navigable annotations.
\newblock In {\em Proc.~of Int. Conf. on Intelligent User Interfaces
  Companion}, IUI '20, page 114–115, 2020.

\bibitem{context_agnostic_engagement}
Sahan Bulathwela, Mar\'{\i}a P\'{e}rez-Ortiz, Aldo Lipani, Emine Yilmaz, and
  John Shawe-Taylor.
\newblock Predicting engagement in video lectures.
\newblock In {\em Proc.~of Int. Conf. on Educational Data Mining}, EDM ’20,
  2020.

\bibitem{peek_orsum}
Sahan Bulathwela, Maria Perez-Ortiz, Erik Novak, Emine Yilmaz, and John
  Shawe-Taylor.
\newblock Peek: A large dataset of learner engagement with educational videos,
  2021.

\bibitem{trueeducation}
Sahan Bulathwela, Mar\'{\i}a P\'{e}rez-Ortiz, Emine Yilmaz, and John
  Shawe-Taylor.
\newblock Towards an integrative educational recommender for lifelong learners.
\newblock In {\em AAAI Conference on Artificial Intelligence}, 2020.

\bibitem{truelearn}
Sahan Bulathwela, Mar\'{\i}a P\'{e}rez-Ortiz, Emine Yilmaz, and John
  Shawe-Taylor.
\newblock Truelearn: A family of bayesian algorithms to match lifelong learners
  to open educational resources.
\newblock In {\em AAAI Conference on Artificial Intelligence}, 2020.

\bibitem{quality_features}
Sahan Bulathwela, Emine Yilmaz, and John Shawe-Taylor.
\newblock Towards automatic, scalable quality assurance in open education.
\newblock In {\em Workshop on AI and the United Nations SDGs at Int. Joint
  Conf. on Artificial Intelligence}, 2019.

\bibitem{choi2020ednet}
Youngduck Choi, Youngnam Lee, Dongmin Shin, Junghyun Cho, Seoyon Park, Seewoo
  Lee, Jineon Baek, Chan Bae, Byungsoo Kim, and Jaewe Heo.
\newblock Ednet: A large-scale hierarchical dataset in education.
\newblock In {\em International Conference on Artificial Intelligence in
  Education}, pages 69--73. Springer, 2020.

\bibitem{Corbett2001}
Albert Corbett.
\newblock Cognitive computer tutors: Solving the two-sigma problem.
\newblock In Mathias Bauer, Piotr~J. Gmytrasiewicz, and Julita Vassileva,
  editors, {\em User Modeling 2001}, pages 137--147. Springer Berlin
  Heidelberg, 2001.

\bibitem{corbett1994knowledge}
Albert~T. Corbett and John~R. Anderson.
\newblock Knowledge tracing: Modeling the acquisition of procedural knowledge.
\newblock {\em User Modeling and User-Adapted Interaction}, 4(4), 1994.

\bibitem{curtis2018online}
Vickie Curtis.
\newblock {\em Online citizen science and the widening of academia: Distributed
  engagement with research and knowledge production}.
\newblock Springer, 2018.

\bibitem{dignum2021role}
Virginia Dignum.
\newblock The role and challenges of education for responsible ai.
\newblock {\em London Review of Education}, 19(1):1--11, 2021.

\bibitem{TVET2018}
Max Ehlers, Robert Schuwer, and Ben Janssen.
\newblock Oer in tvet: Open educational resources for skills development, 2018.

\bibitem{idia_report}
IDIA Working~Group for Artificial~Intelligence.
\newblock Artificial intelligence in international development.
\newblock
  \url{"https://static1.squarespace.com/static/5b156e3bf2e6b10bb0788609/t/5e1f0a37e723f0468c1a77c8/1579092542334/AI+and+international+Development_FNL.pdf"},
  2019.

\bibitem{assist_math1}
Emily~R. Fyfe.
\newblock Providing feedback on computer-based algebra homework in
  middle-school classrooms.
\newblock {\em Comput. Hum. Behav.}, 63(C):568–574, October 2016.

\bibitem{datasheets}
Timnit Gebru, Jamie Morgenstern, Briana Vecchione, Jennifer~Wortman Vaughan,
  Hanna~M. Wallach, Hal~Daum{\'{e}} III, and Kate Crawford.
\newblock Datasheets for datasets.
\newblock {\em CoRR}, abs/1803.09010, 2018.

\bibitem{Guo_vid_prod}
Philip~J. Guo, Juho Kim, and Rob Rubin.
\newblock How video production affects student engagement: An empirical study
  of mooc videos.
\newblock In {\em Proc.~of the First ACM Conf. on Learning @ Scale}, 2014.

\bibitem{emerging_pedagogy}
Christothea Herodotou, Mike Sharples, Mark Gaved, Agnes Kukulska-Hulme, Bart
  Rienties, Eileen Scanlon, and Denise Whitelock.
\newblock Innovative pedagogies of the future: An evidence-based selection.
\newblock {\em Frontiers in Education}, 4:113, 2019.

\bibitem{holloway2019disability}
Catherine Holloway.
\newblock Disability interaction (dix) a manifesto.
\newblock {\em Interactions}, 26(2):44--49, 2019.

\bibitem{holmes2019artificial}
Wayne Holmes, Maya Bialik, and Charles Fadel.
\newblock {\em Artificial Intelligence in Education: Promises and Implications
  for Teaching and Learning}.
\newblock Independently Published, 2019.

\bibitem{holmes2021ai}
Wayne Holmes, Zhang Hui, Fengchun Miao, Huang Ronghuai, et~al.
\newblock {\em AI and education: A guidance for policymakers}.
\newblock UNESCO Publishing, 2021.

\bibitem{gdi_report}
Global Disability~Innovation Hub, University~College London, UNESCO's
  International Research~Centre on~Artificial~Intelligence, European~Disability
  Forum, and Jožef~Stefan Institute.
\newblock Powering inclusion: Ai and at. the findings of an online expert
  roundtable.
\newblock \url{"https://at2030.org/powering-inclusion-ai-and-at"}, 2021.

\bibitem{hylen2021open}
Jan Hyl{\'e}n.
\newblock Open educational resources: Opportunities and challenges.
\newblock 2021.

\bibitem{goal_based_edrec}
Weijie Jiang, Zachary~A. Pardos, and Qiang Wei.
\newblock Goal-based course recommendation.
\newblock In {\em Proceedings of International Conference on Learning Analytics
  \& Knowledge}, 2019.

\bibitem{kreutzercorrect}
Julia Kreutzer, Nathaniel Berger, and Stefan Riezler.
\newblock Correct me if you can: Learning from error corrections and markings.
\newblock In {\em 22nd Annual Conference of the European Association for
  Machine Translation}, page 135, 2020.

\bibitem{kukulska2015mobile}
Agnes Kukulska-Hulme, Mark Gaved, Lucas Paletta, Eileen Scanlon, Ann Jones, and
  Andrew Brasher.
\newblock Mobile incidental learning to support the inclusion of recent
  immigrants.
\newblock {\em Ubiquitous Learning: an international journal}, 7(2):9--21,
  2015.

\bibitem{oer_worldmap}
The OER~World Map.
\newblock The oer world map.
\newblock \url{https://oerworldmap.org/resource}, 2021.

\bibitem{forbes_ai}
Bernard Marr.
\newblock The amazing ways how wikipedia uses artificial intelligence.
\newblock
  \url{"https://www.forbes.com/sites/bernardmarr/2018/08/17/the-amazing-ways-how-wikipedia-uses-artificial-intelligence"},
  2018.

\bibitem{its_press_release}
Dan Meyer.
\newblock Is this press release from 2012 or 1972?
\newblock
  \url{"https://blog.mrmeyer.com/2013/is-this-press-release-from-2012-or-1972"},
  2013.
\newblock Retrieved on 01-05-2021.

\bibitem{molnar2020interpretable}
Christoph Molnar.
\newblock {\em Interpretable machine learning}.
\newblock Lulu. com, 2020.

\bibitem{nilsson2016policy}
M{\aa}ns Nilsson, Dave Griggs, and Martin Visbeck.
\newblock Policy: map the interactions between sustainable development goals.
\newblock {\em Nature News}, 534(7607):320, 2016.

\bibitem{x5gon}
Erik Novak, Jasna Urbančič, and Miha Jenko.
\newblock Preparing multi-modal data for natural language processing.
\newblock In {\em Proc.~of Slovenian KDD Conf. on Data Mining and Data
  Warehouses (SiKDD)}, 2018.

\bibitem{pawlowski2007open}
Jan~M Pawlowski and Volker Zimmermann.
\newblock Open content: a concept for the future of e-learning and knowledge
  management.
\newblock {\em Knowtech 2007, Frankfurt}, 2007.

\bibitem{x5gon_d5_3}
Alex P\'{e}rez, Javier Jorge, and Alfons Juan.
\newblock D5.3 – final report on piloting.
\newblock
  \url{https://www.x5gon.org/wp-content/uploads/2021/01/D5.3_8Dec2020.pdf}.
\newblock Accessed in: 2021-08-02.

\bibitem{10.1145/3397482.3450721}
Maria Perez-Ortiz, Claire Dormann, Yvonne Rogers, Sahan Bulathwela, Stefan
  Kreitmayer, Emine Yilmaz, Richard Noss, and John Shawe-Taylor.
\newblock X5learn: A personalised learning companion at the intersection of ai
  and hci.
\newblock In {\em 26th International Conference on Intelligent User Interfaces
  - Companion}, IUI '21 Companion, page 70–74, New York, NY, USA, 2021.
  Association for Computing Machinery.

\bibitem{ircai_report}
Maria Perez-Ortiz, Erik Novak, Sahan Bulathwela, and John Shawe-Taylor.
\newblock An ai-based learning companion promoting lifelong learning
  opportunities for all.
\newblock
  \url{"https://ircai.org/wp-content/uploads/2021/01/IRCAI_REPORT_01.pdf"},
  2020.

\bibitem{10.1145/2993318.2993332}
Guangyuan Piao and John~G. Breslin.
\newblock Exploring dynamics and semantics of user interests for user modeling
  on twitter for link recommendations.
\newblock In {\em Proceedings of the 12th International Conference on Semantic
  Systems}, SEMANTiCS 2016, page 81–88, New York, NY, USA, 2016. Association
  for Computing Machinery.

\bibitem{ramesh2014learning}
Arti Ramesh, Dan Goldwasser, Bert Huang, Hal Daume~III, and Lise Getoor.
\newblock Learning latent engagement patterns of students in online courses.
\newblock In {\em Proc.~of AAAI Conference on Artificial Intelligence}, 2014.

\bibitem{tomavsev2020ai}
Nenad Toma{\v{s}}ev, Julien Cornebise, Frank Hutter, Shakir Mohamed, Angela
  Picciariello, Bec Connelly, Danielle~CM Belgrave, Daphne Ezer, Fanny~Cachat
  van~der Haert, Frank Mugisha, et~al.
\newblock Ai for social good: unlocking the opportunity for positive impact.
\newblock {\em Nature Communications}, 11(1):1--6, 2020.

\bibitem{toyama2010can}
Kentaro Toyama.
\newblock Can technology end poverty.
\newblock {\em Boston Review}, 36(5):12--29, 2010.

\bibitem{unesco1}
UNESCO.
\newblock Open educational resources (oer).
\newblock \url{https://en.unesco.org/themes/building-knowledge-societies/oer},
  2021.
\newblock Accessed: 2021-04-01.

\bibitem{paris_mlw_report}
UNHRC.
\newblock Design sprint on curriculum alignment in crisis contexts.
\newblock
  \url{"https://drive.google.com/file/d/1Js8KFGaQWp_iej81K4qWdUsNCqsYxCEa/view"},
  2019.

\bibitem{curriculum_report}
UNHRC.
\newblock Hackathon on curriculum alignment: Synthesis and next steps.
\newblock \url{"https://www.unhcr.org/5e81cdd24.pdf"}, 2019.

\bibitem{united2020global}
Scientific United Nations~Educational and Cultural Organization.
\newblock Global education monitoring report 2020: Inclusion and education: All
  means all.
\newblock 2020.

\bibitem{voigt2021open}
Maximilian Voigt.
\newblock Open education and open source for sustainable economic activity.
\newblock {\em {\"O}kologisches Wirtschaften-Fachzeitschrift}, 36(O1):25--27,
  2021.

\bibitem{wiki_wang}
Morten Warncke-Wang, Dan Cosley, and John Riedl.
\newblock Tell me more: An actionable quality model for wikipedia.
\newblock In {\em Proc.~of Int. Symposium on Open Collaboration}, WikiSym '13,
  2013.

\bibitem{woolf2010}
Beverly~Park Woolf.
\newblock {\em Building intelligent interactive tutors: Student-centered
  strategies for revolutionizing e-learning}.
\newblock Morgan Kaufmann, 2010.

\bibitem{wiki_troll}
Ellery Wulczyn, Nithum Thain, and Lucas Dixon.
\newblock Ex machina: Personal attacks seen at scale.
\newblock In {\em Proceedings of the 26th International Conference on World
  Wide Web}, WWW '17, page 1391–1399, Republic and Canton of Geneva, CHE,
  2017. International World Wide Web Conferences Steering Committee.

\bibitem{yu2020mooccube}
Jifan Yu, Gan Luo, Tong Xiao, Qingyang Zhong, Yuquan Wang, Wenzheng Feng, Junyi
  Luo, Chenyu Wang, Lei Hou, Juanzi Li, et~al.
\newblock Mooccube: a large-scale data repository for nlp applications in
  moocs.
\newblock In {\em Proceedings of the 58th Annual Meeting of the Association for
  Computational Linguistics}, pages 3135--3142, 2020.

\bibitem{Yudelson13}
Michael~V. Yudelson, Kenneth~R. Koedinger, and Geoffrey~J. Gordon.
\newblock Individualized bayesian knowledge tracing models.
\newblock In H.~Chad Lane, Kalina Yacef, Jack Mostow, and Philip Pavlik,
  editors, {\em Proc.~of Artificial Intelligence in Education}, 2013.

\end{thebibliography}
}

\end{document}